\documentstyle[12pt]{article}
\def\ni{\noindent}  
\def\yma{$\overline {y}_1\:$}
\def\ymb{$\overline {y}_2\:$}
\def\dt{$\Delta T\:$}

\begin{document}
\title
{\Large{\bf Fractal Analysis 
of Electrical \\ Power Time Series}}

\author{\bf J.R. Sanchez, C.M. Arizmendi\\
\rm \normalsize Departamento de F\'{\i}sica, Facultad de Ingenier\'{\i}a\\
\rm \normalsize Universidad Nacional de Mar del Plata\\
\rm \normalsize Av. J.B. Justo 4302\\
\rm \normalsize 7600 Mar del Plata\\
\rm \normalsize Argentina}
\date{}
\maketitle

\begin{abstract}
Fractal time series has been shown to be self-affine and are characterized
by a roughness exponent $H$. The exponent $H$ is a measure of the
persistence of the fluctuations associated with the time series.
We use a recently introduced method for measuring the roughness exponent,
the mobile averages analysis, to compare  
electrical power demand of two different places, a touristic city 
and a whole country. 
\end{abstract}

\newpage
Over the last years, physicists, biologists and economists have found a new way
of understanding the growth of complexity in nature. Fractals are the way of
seeing order and pattern where formerly only the random and unpredictable had
been observed.
The father of fractals, Benoit Mandelbrot, begun thinking in them studying the 
distribution of large and small incomes in economy in the $60$'s. He noticed 
that an economist's article of faith seemed to be wrong. It was the conviction 
that small, transient changes had nothing in common with large, long term 
changes. Instead of separating tiny changes from grand ones, his picture bound 
them together. He found patterns across every scale. Each particular change 
was random and unpredictable. But the sequence of changes was independent of 
scale: curves of daily changes and monthly changes matched perfectly.

Mandelbrot worked in IBM, and after his study in economy, 
he came upon the problem of noise in telephone lines used to 
transmit information between computers. The transmission noise was well 
known to come in clusters. Periods of errorless communication would be 
followed by periods of errors. Mandelbrot provided a way of describing 
the distribution of errors that predicted exactly the pattern of errors 
observed. His description worked by making deeper and deeper separations 
between periods with errors and periods without errors. But within periods 
of errors (no matter how short) some periods completely clean will be found. 
Mandelbrot was duplicating the Cantor set, created by the $19^{th}$ century 
mathematician Georg Cantor. 

In the {\it fractal way} of looking nature roughness and asymmetry are not just 
accidents on the classic and smooth shapes of Euclidian geometry. Mandelbrot 
has said that ``mountains are not cones and clouds are not spheres''. Fractals 
have been named the {\sl geometry of nature} because they can be found 
everywhere in nature:  mammalian lungs,  trees, and  coastlines, to name just 
a few. 

An english scientist, Lewis Richardson around 1920 checked 
encyclopedias  in Spain, Portugal, Belgium and the Netherlands and 
discovered discrepancies of twenty percent in the lengths of their 
common frontiers. In \cite{hein} the authors measured the coast of Britain 
on a geographical map with different compass settings by counting the number 
of steps along the coast with each setting. The smaller the setting they used, 
the longer the length of the coastline they obtained. If this experiment is 
done to measure the perimeter of a circle, (or any other euclidean shape), the 
length obtained converges with smaller compass settings. In his famous book 
\cite{Mandel} Mandelbrot states that the length of a coastline can never be 
actually measured, because it depends on the length of the ruler we use.

Fractal sets show self similarity with respect to space. Fractal time series 
have statistical self similarity with respect to time. In the social and 
economic fields, time series are very common. In \cite{pete} a simple way 
to demonstrate self-similarity in a time series of stock returns is devised 
by asking the reader to guess which graph corresponds to daily, weekly  and 
monthly returns between three different graphs with no scale on the axes. 

An important statistics used to characterize time series is the 
{\it Hurst exponent} \cite{pete}. Hurst was a hydrologist who worked on the 
Nile River Dam project in the first decades of this century. At that time, 
it was common to assume that the uncontrollable influx of water from rainfall 
followed a random walk, in which each step is random and independent from 
previous ones. The random walk is based on the fundamental concept of Brownian 
motion. Brownian motion refers to the erratic displacements of small solid 
particles suspended in a liquid. The botanist Robert Brown, about 1828, 
realized that the motion of the particles is due to light collisions with the 
molecules of the liquid.

Hurst measured how the reservoir level fluctuated around its average level over time. 
The range of this fluctuation depends on the length of time used for measurement. If 
the series were produced by a random walk, the range would increase with the square 
root of time as $T^{1/2}$. Hurst found that the random walk assumption was wrong for 
the fluctuations of the reservoir level as well as for most natural phenomena, like 
temperatures, rainfall and sunspots. The fluctuations for all this phenomena may be 
characterized as a ``biased random walk''-a trend with noise- with range increasing 
as $T^H$, with $H > 0.5$. Mandelbrot called this kind of generalized random walk 
{\sl fractional brownian motion}. In high-school statistical courses we have been 
taught that nature follows the gaussian distribution which corresponds to random 
walk and $H=1/2$. Hurst's findings show that it is wrong.

The proper range for $H$ is from $0$, corresponding to very rough random fractal 
curves, to 1 corresponding to rather smooth looking fractals. In fact, there is a 
relation between $H$ and the fractal dimension $D$ of the graph of a random fractal: 
\begin{equation}
\label{eq1}
D=2-H.
\end{equation} 
Thus, when the exponent $H$ vary from $0$ to $1$, yields dimensions 
$D$ decreasing from $2$ to $1$, which correspond to more or less wiggly 
lines drawn in two dimensions.

Fractional Brownian motion can be divided into three distinct categories: 
$H<1/2$, $H=1/2$ and $H>1/2$. The case $H=1/2$ is the ordinary random walk or 
Brownian motion with independent increments which correspond to normal 
distribution.

For $H<1/2$ there is a negative correlation between the increments. This type 
of system is called {\it antipersistent}. If the system has been up in some 
period, it is more likely to be down in the next period. Conversely, if it was 
down before, it is more likely to be up next. The antipersistence strength 
depends on how far $H$ is from $1/2$.

For $H>1/2$ there is a positive correlation between the increments. 
This is a {\it persistent} series. If the system has been up (down) 
in the last period, it will likely continue positive (negative) in the 
next period. Trends are characteristics of persistent series. The strength 
of persistence increases as $H$ approaches $1$. Persistent time series are 
plentiful in nature and in social systems. As an example the Hurst exponent 
of the Nile river is $0.9$, a long range persistence that requires unusually 
high barriers, such as the Asw\^an High Dam to contain damage in the floods.

Generally, experimental time series don't have a unique $H$. In that case time 
series may be considered as a {\sl multifractal}\cite{hein}, $H$ is associated with 
the {\sl singularity exponent $\alpha$} restricted to an interval
$\alpha_{min}<\alpha<\alpha_{max}$ characterized  by 
the {\sl singularity spectrum $f(\alpha)$}.
 
In order to analyze the fractal properties of electrical power consumption, 
we used a recently developed technique to obtain Hurst exponents 
of self-affine time series based on the so-called mobile averages \cite{vande}.  
The method is introduced as follows: consider a time series $y(t)$ given at 
discrete times $t$, a mobile average 
$\overline {y}(t)$ is defined as
\begin{equation}
\overline {y}(t) = \frac{1}{T}\sum_{i=0}^{T-1} y(t-i)\:,
\end{equation}
\ni i.e., the average of the last $T$ data points. 
Its can be easily show that if $y(t)$ increases (decreases) with time, 
$\overline {y} < y$ ($\overline {y} > y$). Thus the mobile average captures the
trend of the signal over the time interval $T$. Although the mobile average is
used mainly for studying the behavior of financial time series, the procedure 
can be, in fact, used on any times series.

Now, consider two different mobile
averages \yma and \ymb characterized over intervals $T_1$ and $T_2$ respectively,
such that $T_2 > T_1$. As it is well known in stock chart analysis, the crossing of
\yma and \ymb coincide with drastic changes of the trend of $y(t)$. If $y(t)$ 
increases for a long period before decreasing rapidly, \yma will cross \ymb from
above. On the contrary, if \yma crosses \ymb from below, the crossing point
coincides with an upsurge of the signal $y(t)$. This type of crosses are used in
empirical finance to {\it extrapolate} the evolution of the market.

If a signal is self-affine, it shows scaling properties of the form
\begin{equation}
y(t) \sim b^{-H}\:y(bt)\:,
\end{equation}
\ni where the exponent $H$ is the Hurst exponent. 

It is well known \cite{feder} that the set of crossing points between the signal $y(t)$
and the $y=0$ level is a Cantor set with fractal dimension $1-H$. On the other
hand, in reference \cite{vande} the question of whether there is a Cantor set for the
crossing points between \yma and \ymb is raised. The density $\rho$ of such
crossing points is calculated for various artificially generated time series with
different values of $H$. In all the checked cases, $\rho$ is independent of the
size $N$ of the time series and the fractal dimension of the set of crossing
points is $1$, i.e., the points are homogeneously distributed in time along \yma
and \ymb. 
Due to the homogeneous distribution of the crossing points, the
forecasting of the crossing points between \yma and \ymb is impossible even for
self-affine signals $y(t)$. 
However, in the same reference \cite{vande}, it is shown that there
exists a scaling relation between $\rho$ and \dt, where \dt is defined as
\begin{equation}
\Delta T = \frac{T_2 - T_1}{T_2}\:.
\end{equation}
\ni With this definition, it is shown that $\rho$ scales as 
\begin{equation}
\label{eq5}
\rho \sim \frac{1}{T_2}\left[(\Delta T)(1-\Delta T)\right]^{H-1}\:,
\end{equation}
\ni i.e., for small values of \dt, $\rho$  scales as $\Delta T^{H-1}$, while
the relation $\rho \sim (1-\Delta T)^{H-1}$ is valid for \dt close to $1$.

One practical interest of the above relations stems in the easy implementation
of an algorithm for measuring $H$. 
We have used the scaling properties of $\rho$ in order to estimate the fractal 
exponent for two time series representing the electrical power demand of two completely 
different places: Australia, a whole continent, and Mar del Plata a touristic city 
of Argentina. Data of Australia electrical power demand were obtained  at the web site:
http://www.tg.nsw.gov.au/sem/ \\ statistics, while
Mar del Plata electrical demand time series was kindly
provided by Centro Operativo de Distribucion Mar del Plata belonging to
EDEA, a local energy distribution enterprise. 

Two time series of 8832 points were taken, as seen in Fig. 1 a) and b). 
In Fig. 2, the density $\rho$ of crossing points as a function of the relative 
difference $\Delta T$, with $T_2 = 100$, is shown for Australian time
series. The slope through the point is $0.264 \pm 0.024$, which means 
$H \cong 0.736$ according to equation \ref{eq5}. 
In Fig. 3, the density $\rho$ of crossing points for the same value of $T_2$
is shown for Mar del Plata time series. In this case, the corresponding 
slope through the point is $0.218 \pm 0.012$, which means $H \cong 0.782$.
From the exponent $H$, the corresponding fractal dimensions $D_f$ can be calculated
using equation \ref{eq1}.
In both cases similar exponents $H$ are obtained with approximately the same
degree of persistence, although both time series correspond to completely
different kind of electrical power consumption, a continent and a touristic city.
A more detailed fractal analysis of these time series was recently done by one of us 
\cite{iee} using wavelets.
The wavelet technique of fractal analysis is more powerful than the mobile
averages technique, but much more complicated. In \cite{iee} both time series
were characterized as multifractals.
However, the results reported here for the roughness exponent have a good
agreement with the value $H$ associated with the maximum of the multifractal
spectrum.

\newpage
FIGURE CAPTIONS

\vspace{0.4cm}
Fig. 1. a) Electrical power demand time series for Australia. b) 
Electrical power demand time series for Mar del Plata.

\vspace{0.4cm}
Fig. 2 Log-log plot of the density $\rho$ of crossing points as a 
function of the relative 
difference $\Delta T$, with $T_2 = 100$, for Australian time
series. The slope through the point is $0.264 \pm 0.024$, which means 
$H \cong 0.736$.

\vspace{0.4cm}
Fig. 3 Log-log plot of the density $\rho$ of crossing points as a 
function of the relative 
difference $\Delta T$, with $T_2 = 100$, for Mar del Plata time series. 
The corresponding slope through the point is $0.218 \pm 0.012$, 
which means $H \cong 0.782$.

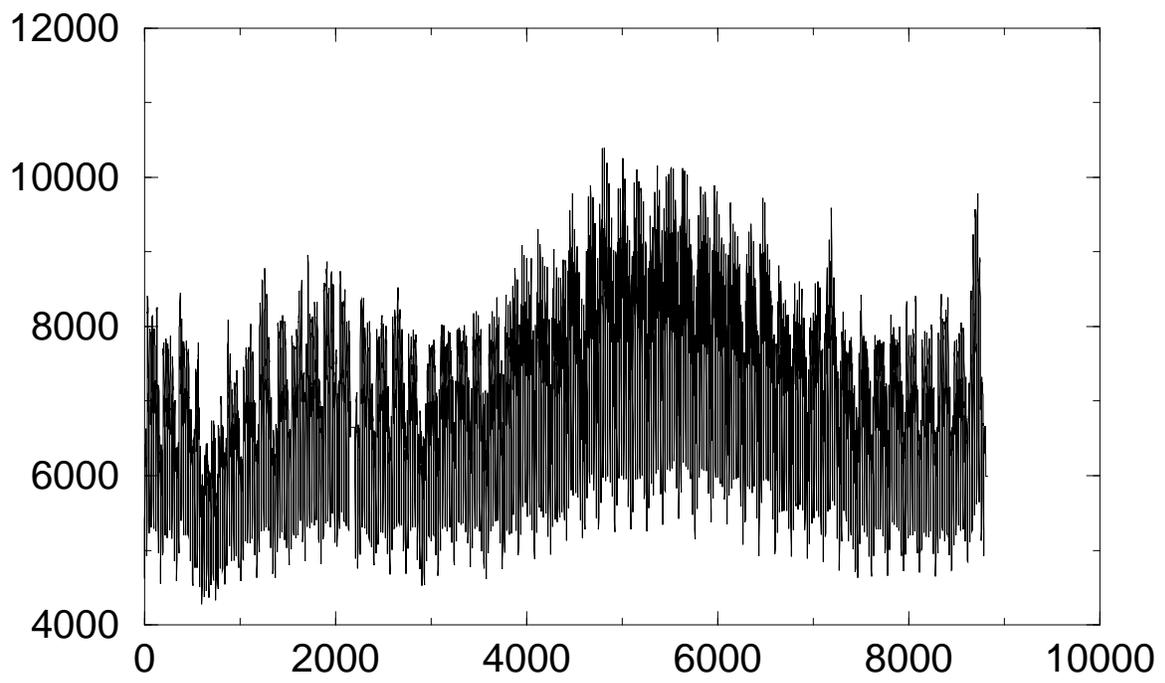
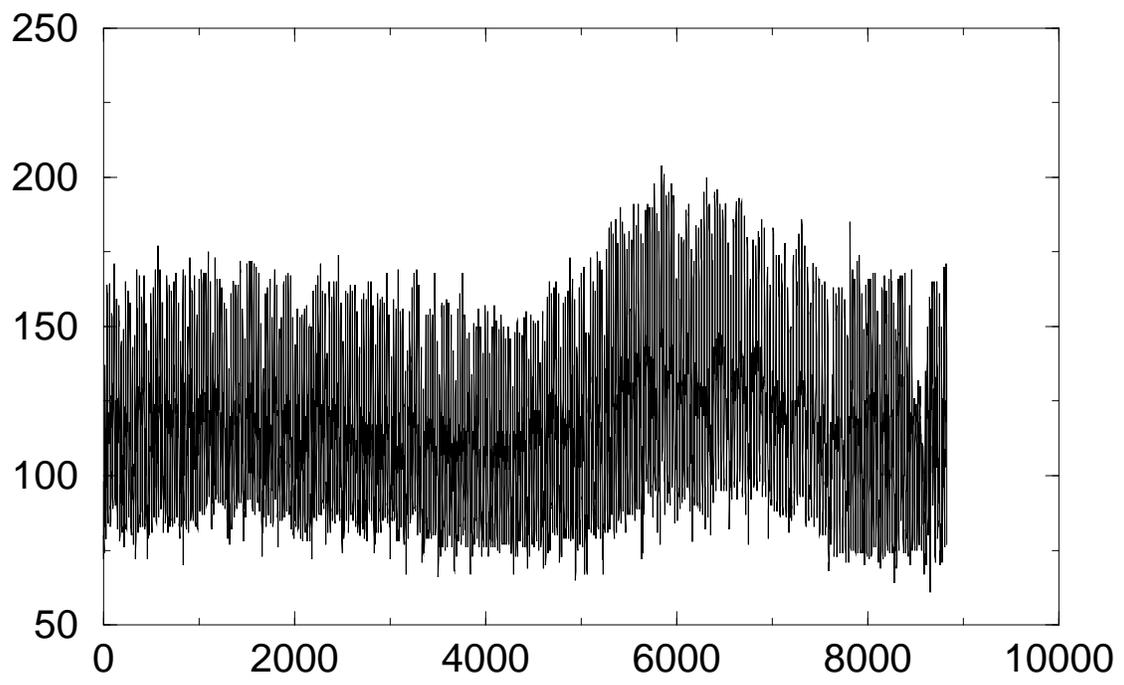
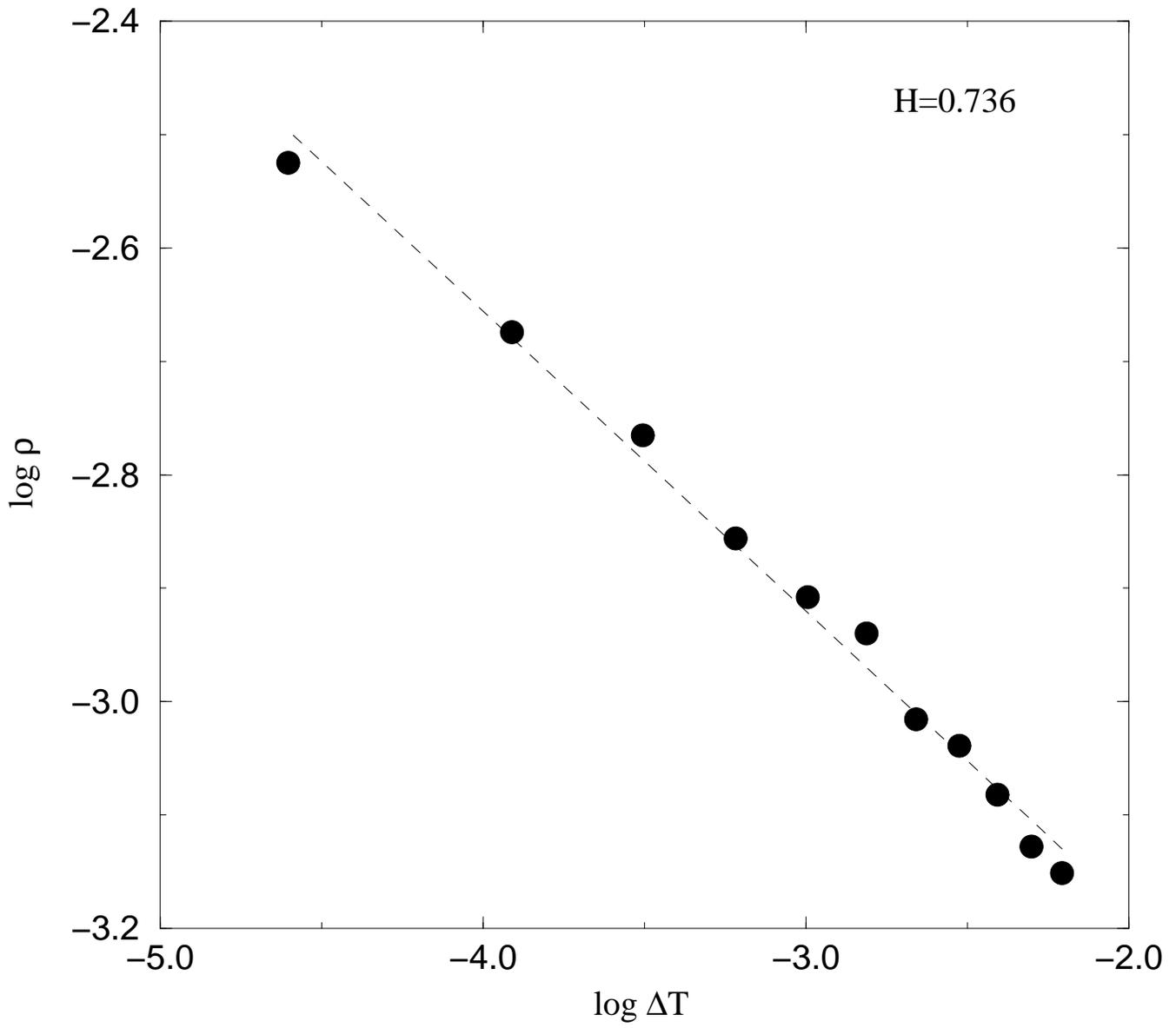
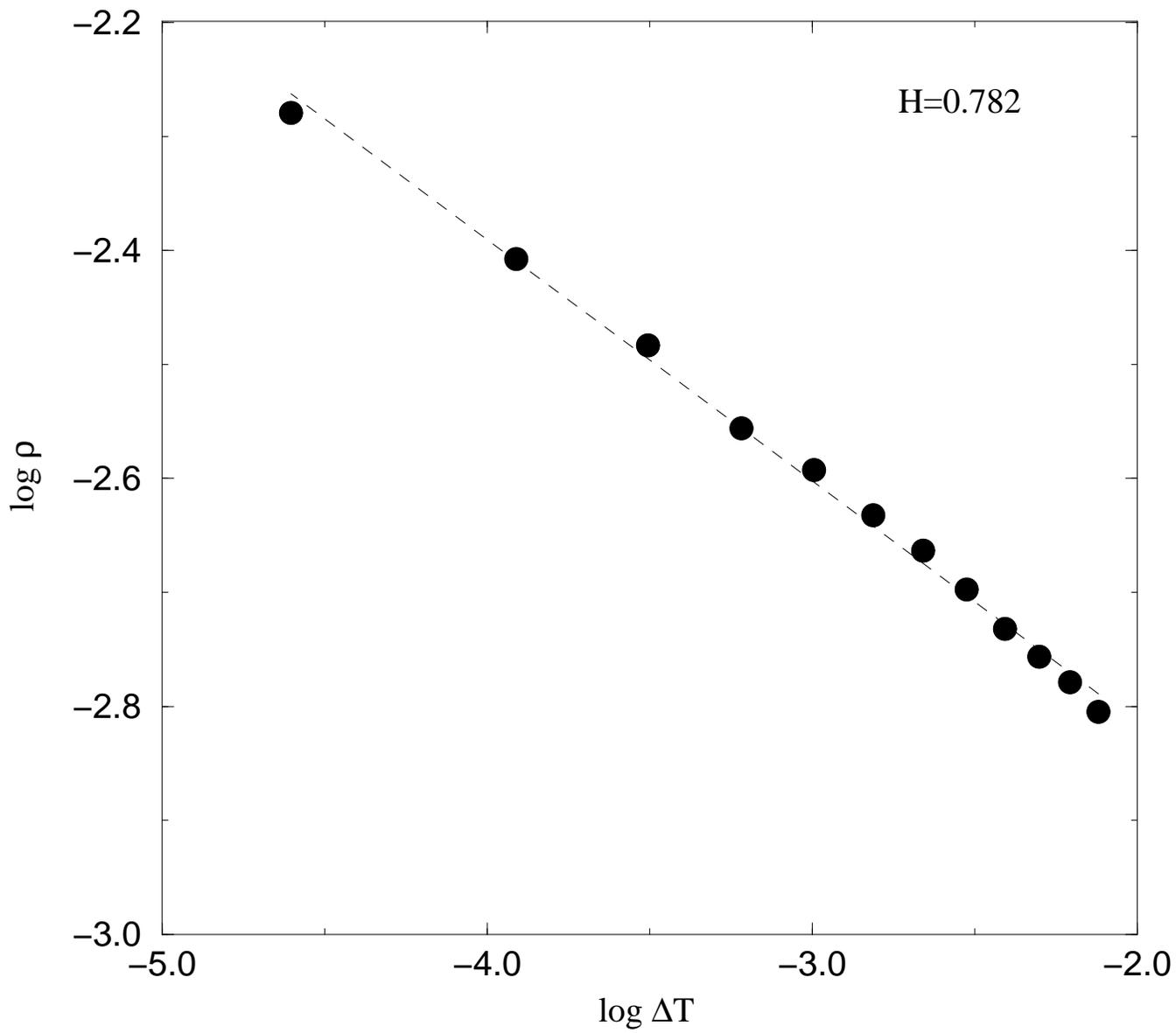
\end{document}